\documentclass[12pt, hidelinks]{article}
\usepackage[small, bf]{caption}
\usepackage[utf8]{inputenc}
\usepackage{fullpage}
\usepackage{appendix}
\usepackage{graphicx}
\usepackage{amsmath}
\usepackage{amssymb}
\usepackage{amsthm, amsfonts}

\newcommand{\downto}{\downarrow}

\newcommand{\ones}{\mathbf 1}
\newcommand{\reals}{{\mbox{\bf R}}}

\newcommand{\Expect}{\mathop{\bf E{}}}

\newcommand{\eg}{{\it e.g.}}
\newcommand{\ie}{{\it i.e.}}

\newcommand{\BEAS}{\begin{eqnarray*}}
\newcommand{\EEAS}{\end{eqnarray*}}
\newcommand{\BEA}{\begin{eqnarray}}
\newcommand{\EEA}{\end{eqnarray}}
\newcommand{\BEQ}{\begin{equation}}
\newcommand{\EEQ}{\end{equation}}
\newcommand{\BIT}{\begin{itemize}}
\newcommand{\EIT}{\end{itemize}}

\title{Replicating Market Makers}

\author{
Guillermo Angeris\\
{\small \texttt{angeris@stanford.edu}}
\and
Alex Evans\\
{\small \texttt{alex@placeholder.vc}}
\and
Tarun Chitra\\
{\small \texttt{tarun@gauntlet.network}}}

\date{March 2021}

\begin{document}

\maketitle

\begin{abstract}
We present a method for constructing Constant Function Market Makers (CFMMs) whose portfolio value functions match a desired payoff.
More specifically, we show that the space of concave, nonnegative, nondecreasing, 1-homogeneous payoff functions and the space of convex CFMMs are equivalent; in other words, every CFMM has a concave, nonnegative, nondecreasing, 1-homogeneous payoff function, and every payoff function with these properties has a corresponding convex CFMM. We demonstrate a simple method for recovering a CFMM trading function that produces this desired payoff.
This method uses only basic tools from convex analysis and is intimately related to Fenchel conjugacy. We demonstrate our result by constructing trading functions corresponding to basic payoffs, as well as standard financial derivatives such as options and swaps.

\end{abstract}

\section*{Introduction}

Constant Function Market Makers (CFMMs)~\cite{AC20} are a family of automated market makers that enable censorship-resistant asset exchange on public blockchains.
CFMMs are capitalized by liquidity providers (LPs) who supply reserves to an on-chain smart contract. The CFMM uses these reserves to execute swaps for traders, allowing a swap only if it preserves some function of reserves, known as the trading function or invariant.
For example, Uniswap~\cite{uniswap} only allows trades that keep the product of the reserves after the trade equal to the product of the reserves before the trade.

A key question first explored in~\cite{angeris2019analysis} concerns the returns that LPs receive for their capital.
As shown in~\cite{AC20}, the value of the LP's assets in a CFMM can be determined by solving a convex problem over the CFMM's trading set.
This method allows one to compute explicit expressions for the value LPs receive from most popular CFMMs, such as Uniswap and Balancer, as well practical lower bounds for a larger class for trading functions.
Additional work such as \cite{AEC20,AEC21} has explored the impact of parameters such as trading-function curvature and swap fees on LP returns.

Here, we consider what may be called the `inverse' problem.
Rather than deriving the LP's payoff for a given trading function, we seek to find the trading function that guarantees that LPs receive a certain payoff.
By contributing capital to the CFMM, LPs will statically replicate their desired payoff.
This is a generalization of the problem considered in~\cite[\S5]{evans2020liquidity} that 
shows how to replicate continuously-differentiable payoffs using constant mean trading 
functions with dynamically-adjusted weights.
These constructions require continual updates from on-chain oracles that may be expensive, 
complex to manage, and are often vulnerable to front-running attacks.
In contrast, the trading functions we derive in this work are not time-varying and do not 
depend on external price oracles, so they are likely easier to implement in practice. 

\paragraph{Hedging.} One obvious question is: why would one want to implement such desired payoffs?
Trading on blockchains has a number of idiosyncrasies that make dynamic hedging strategies 
expensive to execute. In particular, unlike centralized venues, most blockchains prevent spam 
and denial of service attacks by charging users a fee per transaction. This fee, known as gas 
on networks like Ethereum, can be a dominant cost to on-chain traders during times of high 
market volatility~\cite{daianFlashBoysFrontrunning2019,kao2020analysis}. Moreover, the 
volatility of gas costs on Ethereum makes on-chain dynamic hedging strategies more complex to 
manage successfully. 

In contrast, CFMMs allow the LP to achieve a desired payoff by passively contributing capital.
Rather than requiring LPs to continually rebalance their holdings through on-chain trades, CFMMs incentivize arbitrageurs to adjust reserves to the level required to achieve their desired hedge.
In effect, this approach outsources the cost and complexity of on-chain trading to specialized parties.
As discussed in~\S\ref{app:delta-hedging}, the added simplicity for the hedger is not without trade-offs as LPs are subject to arbitrage losses. 
However, small fees have been shown mitigate these costs in certain settings~\cite{AEC21}. 

\paragraph{Limitations.} The payoffs one can replicate using the methodology presented in this 
paper are limited to concave, nonnegative, nondecreasing functions of price. An instructive analogy is as 
follows. In limit order books, a resting limit order can roughly be thought of as an option which the market 
maker sells to participants---who are executing market (liquidity removing) orders---that allows them to purchase 
or sell a quantity of an asset at a given price~\cite[Chapter~6]{trading_handbook}. (The 
precise replication of a market maker's portfolio of limit orders as a covered options is 
complicated somewhat by the fact that trades depend on the positions in the 
queue~\cite{foucault2007does,moallemi2014value}, but this point is not essential here.) 
Being short an option generates a concave, or `negative gamma,' payoff. By virtue of 
allowing users to execute a set of trades at predetermined prices, CFMMs can also be seen 
as having negative gamma payoffs which lead to `impermanent loss'~\cite{AC20, AEC20, 
clark2020replicating} for LPs. On the other hand, replicating convex payoffs (such as long 
positions in options) requires the ability to short shares in a CFMM, or to use external 
price oracles, as described in~\cite{evans2020liquidity}.

\paragraph{Summary.} The outline of this article is as follows.
In the next section, we describe the problem of constructing a trading function that produces a CFMM with a desired payoff function.
In \S\ref{sec:general_soln}, we present some basic definitions along with a solution method for recovering the desired trading set and corresponding (equivalent) trading functions.
We derive trading functions for some basic examples such as linear and quadratic payoffs in \S\ref{sec:basic-ex}.
In \S\ref{sec:practical}, we proceed with practical applications, such as recovering the constant-mean function used in Balancer~\cite{balancer} as well as constructing trading functions for replicating the Black-Scholes prices of covered European calls and perpetual American puts. We give some possible future directions in~\S\ref{sec:conclusion}.

\section{General solution}\label{sec:general_soln}
In this section, we will present a general method for constructing a CFMM trading function whose value function
matches a desired payoff function, within a reasonable domain. We start with some basic definitions
and provide a relatively general solution method. We continue with some basic (and not so basic)
applications of the method.

\paragraph{Trading function.} A (path independent) CFMM is defined by its \emph{trading 
function} $\psi: \reals^n \to \reals$ and its \emph{reserves} $R \in \reals^n_+$. The
reserve $R_i$ specifies the quantity of coin $i$ available to the CFMM contract, while
the function $\psi$ specifies the behavior of the contract. More specifically, the
contract will allow any agent to trade with the reserves, so long as the
new reserves, $R' \in \reals_+^n$, after the agent has added or withdrawn the required
quantities, satisfy
\[
\psi(R') \ge \psi(R).
\]
Some definitions require that the inequality be an exact equality, but this point
is not essential, since, in practice $\psi$ can always be made an increasing function in its 
arguments, so any rational agent will ensure that the inequality is saturated; see, 
\eg,~\cite[\S2.1]{AC20} for more.

\paragraph{Portfolio value function.} We will assume there exists some external market
with a fixed reference price $c \in \reals_+^n$. Here $c$ is the \emph{price vector} for the 
$n$ coins the CFMM trades, such that $c_i$ is the price of coin $i$ in this external market. 
We will call the total value of reserves, after arbitrage, the \emph{portfolio value} or 
\emph{liquidity provider payoff} of a CFMM, represented by some function $V: \reals_+^n\to 
\reals$. In the case where the CFMM is path-independent, with concave nonincreasing trading 
function $\psi: \reals^n \to \reals$ and constant $k$,~\cite[\S2.5]{AC20} shows that the 
function $V$ is equal to
\[
V(c) = \inf \{c^TR \mid \psi(R) \ge k, \, R \in \reals_+^n\}.
\]
The economic interpretation of this definition of $V$ is simple: an arbitrageur is engaged in 
a zero-sum game with liquidity providers. The arbitrageur's payoff is maximized when the 
portfolio value, $c^TR$, of a liquidity provider is minimized over the valid reserves $R$; \ie,
those that satisfy $\psi(R) \ge k$. (This is a simple restatement of the
optimal arbitrage problem. For more see, \eg,~\cite{AC20, AEC21}.)

Very generally, we will define the portfolio value over a feasible set of reserves $S\subseteq \reals^n$ as
\begin{equation}\label{eq:pv}
V(c) = \inf \{c^TR \mid R \in S\}.
\end{equation}
This includes the previous definition by setting $S = \{R \in \reals^n_+ \mid \psi(R) \ge k\}$,
and this `more general' definition will slightly simplify the derivation presented below. (In fact,
both definitions are equivalent in that, given a convex set $S$, we can construct a concave function $\psi$ whose 0-superlevel set is equal to $S$. We give an explicit construction later in this section.)


\paragraph{Desired payoff.} One very natural question is: given a desired payoff function (\ie, a desired $V$), is it possible
to create a trading function $\psi$ which results in this payoff? Another slightly more casual way of phrasing this problem is: can we
`invert' formula~\eqref{eq:pv}, given $V$?

In general, the answer is no. It is nearly immediate, given any
$\psi$, the function $V$ is always a concave function because it is the infimum of a family of 
linear functions, indexed by $R \in S$.
Second, because $V(c)$ is some infimum over $c^TR$ with both $c$ and $R$ nonnegative, it 
must be nonnegative. Third, given any $c' \ge c \ge 0$, we have that
\[
V(c') = V(c + c' - c) \ge V(c)+ V(c' - c) \ge V(c),
\]
so $V$ is nondecreasing in its arguments.
And, finally, note that the function $V$ is 1-homogeneous in terms of $c$, by definition; \ie, for any $\eta \ge 0$, the payoff function $V$ satisfies
\[
V(\eta c) = \eta V(c).
\]
This limits the set of payoff functions we can find a CFMM trading function 
for, since $V$ must be concave, nonnegative, nondecreasing, and 1-homogeneous in order for there
to exist a CFMM trading function with $V$ as its liquidity provider payoff. (We will
see that 1-homogeneity is not as strong of a condition as it appears at first glance,
and we show how to deal with this in~\S\ref{sec:basic-ex}.) 

\paragraph{Consistent payoff functions.} We will then say $V$ is a \emph{consistent payoff function} if it is concave, nonnegative, nondecreasing, and 1-homogeneous. Clearly, then, the value function of any CFMM will always be consistent, from the discussion above. In the next section, we will also see that the converse of the above statement is true: any consistent payoff function has a path independent CFMM that yields this payoff. We will also show how to construct a CFMM with a desired consistent payoff.

\paragraph{Discussion.} The conditions above all have nice economic interpretations. The concavity of the function implies that `impermanent loss', 
also known as `negative gamma' in finance, is an intrinsic property of liquidity provision in 
path-independent CFMMs, as it holds for any possible CFMM in practice. The nonnegativity 
simply implies that a liquidity provider position always has nonnegative value, while the fact 
that $V$ is nondecreasing implies that the position of liquidity providers does not get worse 
as coins increase in value. Finally, the 1-homogeneity is a notion of `scale-invariance,' \ie, 
scaling the num\'eraire should simply scale the total portfolio value of a liquidity 
provider's position. While we have shown that these conditions are necessary, we will also 
show that they are sufficient in the remainder of this section.

\subsection{Solution method and equivalence} \label{sec:solution-method}
The method presented here can be seen as a special case of Fenchel conjugacy, with some slight modifications.
Though not necessary, since we will introduce the tools required in this note and give self-contained proofs, the results here are essentially corollaries of well-known theorems in convex analysis, with the most notable being strong duality. We refer the reader to, \eg,~\cite[\S5]{cvxbook} for further reading.

Given a consistent payoff function $V$, we will first find a set of reserves $S$ corresponding to the payoff function $V$.
We will then show that the set $S$ will also have $V$ as its payoff function, as defined in~\eqref{eq:pv}. We then find an explicit trading function, $\psi_V$, whose 0-superlevel set is equal to $S$ and therefore has the desired payoff function, $V$.

\paragraph{Feasible reserve set.} Given some concave payoff function $V: \reals^n_+ \to \reals$,
we will define its \emph{feasible reserve set} $S \subseteq \reals^n_+$ as
\begin{equation}\label{eq:feasible-reserve-set}
S = \{R \in \reals^n_+ \mid V(c) \le c^TR, ~\text{for all} ~ c \in \reals^n\}.
\end{equation}
In other words, $S$ is the set of reserves for which the portfolio value of the reserves, at any cost vector $c$, is always no smaller than $V(c)$. Note that the set $S$
is convex as it is the intersection of a family of hyperplanes parametrized by $c$.


\subsubsection{Equivalence}\label{sec:equivalence}

We will show that, if $V$ is a nonnegative, concave, 1-homogeneous
payoff function with feasible reserve set $S$, then $S$ has payoff, as defined in~\eqref{eq:pv}, given by $V$.

Using the definition of $S$ in~\eqref{eq:feasible-reserve-set}, we can rewrite 
problem~\eqref{eq:pv} in extended form 
\begin{equation}\label{eq:payoff}
\begin{aligned}
	& \text{minimize} && c^TR\\
	& \text{subject to} && V(q) \le q^TR, \quad \text{for all} ~ q \in \reals^n,
\end{aligned}
\end{equation}
with variable $R \in \reals^n$. We will write $V^\star(c)$ for the optimal value of~\eqref{eq:payoff}, which depends on $c$.

\paragraph{Lower bound.} Clearly, we know that
\[
V^\star(c) \ge V(c),
\]
since, if $R^\star$ is optimal for $c$, then
\[
V^\star(c) = c^TR^\star \ge V(c),
\]
where the inequality follows from the fact that $R^\star$ is a feasible point for problem~\eqref{eq:payoff}. (In the case that an optimal value doesn't exist, though one can show it always does, we may replace $R^\star$ with a sequence that is feasible and converges to the optimal value.)

\paragraph{Upper bound.} 
It will suffice to show that there exist some reserves $R \in S$ 
such that $c^TR = V(c)$; \ie, $R$ is feasible for problem~\eqref{eq:payoff} with objective
value equal to $V(c)$. Because $R$ is feasible for~\eqref{eq:payoff}, then, by definition
of optimality, we will have that $c^TR \ge V^\star(c)$, as required.

First, pick any $R \in \partial V(c)$; \ie, $R$ is a supergradient of $V$ at $c$, 
such that, for any $q \in \reals^n$ we have
\begin{equation}\label{eq:supergrad}
V(q) \le V(c) + R^T(q - c).
\end{equation}
Such an $R$ exists because $V$ is a concave function and is nonnegative because $V$ is nondecreasing. (As a side note, $R$ is equal to $\nabla V(c)$ when 
the function $V$ is differentiable at $c$. See, \eg,~\cite[Thm.\ 25.1]{rockafellar1970convex}.) Now, let $q = 0$ in~\eqref{eq:supergrad} to get
\[
c^TR \le V(c),
\]
where $V(0) = 0$ by the homogeneity of $V$.
On the other hand, let $q = 2c$ in~\eqref{eq:supergrad} to get
\[
V(2 c) \le V(c) + c^TR.
\]
Since $V(2c) = 2V(c)$ because $V$ is 1-homogeneous, we have
\[
V(c) \le c^TR,
\]
so $V(c) = c^TR$. We can now rearrange~\eqref{eq:supergrad} to get
\[
V(q) - q^TR \le V(c) - c^TR = 0,
\]
for any $q$, which means that $R \in S$ by the definition of $S$
in~\eqref{eq:feasible-reserve-set}.
From before, this means that $R$ is feasible for~\eqref{eq:payoff} and, because
its objective value is $c^TR = V(c)$, we must have that $V(c) \ge V^\star(c)$.

\paragraph{Result.} Putting both statements together yields that $V(c) = V^\star(c)$ for every
$c$, and therefore the set $S$ has the desired payoff. In fact, the proof given above has a
few simple but important consequences. For example, dropping the monotonicity requirement
on $V$ implies that there might exist reserves in $S$ which are negative. One way to deal
with such a problem is to take the intersection of $S$ with the nonnegative reals, $S \cap 
\reals_+^n$, but then $V$ need not equal $V^\star$ as defined above, and the proof above
also gives a simple way of quantifying the gap. Similar results and extensions also
hold for the nonnegativity requirement for $V$, the concavity of $V$, and so on,
which we leave as open questions for future research.

\subsubsection{Constructing a trading function}
From the previous discussion, given a
desired payoff function $V$ we can easily find a trading set $S$ such that $S$ has
payoff equal to $V$. While this may suffice for some applications, it is often
easier to work with the functional form of a CFMM. More specifically, we will look
for a concave trading function $\psi_S$ such that some superlevel set of the function
is equal to the set $S$.

One example of such a functional form (there are many equivalent ones) is given by
\begin{equation}\label{eq:main}
\psi_V(R) = \inf_c (c^TR - V(c)),
\end{equation}
where $c \in \reals^n$ ranges over all real $n$-vectors.
This function has the desired property since
\[
\psi_V(R) \ge 0, ~~ \text{if, and only if}, ~ c^TR \ge V(c)~ \text{for all} ~ c \in \reals^n,
\]
and therefore $\psi_V(R) \ge 0$ if, and only if, $R \in S$,
so $\psi_V$ can be used as the CFMM trading function, as required. We also note that $\psi_V$ is the
negative Fenchel conjugate of $-V$, with negated arguments, \ie:
\[
\psi_V(R) = -\sup_c (-c^TR - (-V)(c)) = -(-V)^*(-R).
\]
This equation, along with the portfolio value equation in~\cite[\S2.5]{AC20} implies, roughly speaking, that the portfolio value function
and the trading function for a CFMM are essentially Fenchel conjugates of each other.

\paragraph{Discussion.} In general, unless certain conditions are satisfied, it is possible
that applying equation~\eqref{eq:main} directly to any desired payoff function $V$ need not
yield a CFMM whose payoff function is equal to $V$ at all prices. We only guarantee equality 
in the case that $V$ is consistent, but find that this procedure is also useful in cases
where $V$ is not. (As discussed, the proof given in~\S\ref{sec:equivalence} gives a way of 
quantifying how much the payoff might differ in cases where $V$ is not consistent.) 
Additionally, we note that the function~\eqref{eq:main} will always be a set-indicator
function; \ie, $\psi_V$ will always be either 0 or $+\infty$ at every point, due to the
1-homogeneity of $V$. In some cases, the set-indicator description can be simplified,
but this need not always be true.


\section{Basic examples and properties} \label{sec:basic-ex}
We show two basic examples, where the desired payoff is linear
and, later, quadratic, and introduce some basic tools which help
simplify derivations.

\subsection{Linear payoffs and offsets}
In this case, we will find a CFMM that produces a linear payoff function; \ie,
what is a CFMM that corresponds to the payoff function:
\[
V(c) = a^Tc,
\]
where $c \in \reals_+^{n}$, $a \in \reals_+^{n}$. It is not difficult to
intuit what the behavior of the LP (and therefore, of the CFMM's trading function) should be. In
order to replicate this payoff, the LP should simply hold $a_i$ of asset $i$, which would
be equivalent to the CFMM disallowing any trades other than the null trade. We will apply the method to 
this example, where the solution is known, as a simple, but potentially useful exercise.

\paragraph{Linear payoff.} As before, we have that
\[
\psi_V(R) = \inf_c(c^TR - a^Tc) = \inf_c ((R - a)^Tc) = \begin{cases}
	0 & R = a\\
	-\infty & \text{otherwise},
\end{cases}
\]
which is exactly the CFMM we expect from the intuitive result: the CFMM can only allow trading if it 
a trade leaves the reserves at $a$.


\paragraph{Linear offsets.} A useful and general tool used in the previous derivation is
that a linear offset of the payoff function $V$ results in a linear offset of the
arguments of the trading function. More specifically, given a payoff function $V$, with trading
function $\psi_V$, the trading function corresponding to the `linearly offset'
payoff:
\begin{equation}\label{eq:linear-offset}
V'(c) = V(c) + a^Tc
\end{equation}
where $a \in \reals_+^n$, is
\[
\psi_{V'}(R) = \psi_V(R - a).
\]
This follows immediately from~\eqref{eq:main} and has an obvious economic interpretation: any linear 
offset in the value function is simply equivalent to adding that quantity of coins to the reserves.
This may help in simplifying derivations since a number of payoffs are simply linear offsets
of other, potentially well-known payoff functions.

\subsection{Perspective transform and quadratic payoffs}
It is also sometimes easier in practice to specify the payoff
with respect to a num\'eraire, rather than with respect to a general price vector. For example, if we
assume that the $n$th coin is the num\'eraire, and $c'$ is the price vector for the first $n-1$ coins
with respect to the $n$th coin, then this is equivalent to specifying the `reduced payoff function'
$U(c')$ for each $c' \in \reals^{n-1}$, which depends only on the first $n-1$ coins.

\paragraph{Perspective transform.} A simple approach
to constructing an $n$ coin, 1-homogeneous payoff function that is concave, nonnegative whenever $U$ 
is also concave, nonnegative is by the use of the \emph{perspective transform} of $U$, which
we define here as the function $V : \reals^n \to \reals$ such that
\begin{equation}\label{eq:perspective}
V(c', c_n) = \begin{cases}
	c_n U(c'/c_n) & c_n > 0\\
	-\infty & \text{otherwise},
\end{cases}
\end{equation}
where $c' \in \reals^{n-1}_+$ is the price of the first $n-1$ coins while $c_n \in \reals_+$
is the price of the num\'eraire. The concavity of $V$, given the concavity of $U$, follows from a basic 
argument (see, \eg,~\cite[\S3.2.6]{cvxbook}), while positivity
is immediate from the definition. The fact that $V$ is 1-homogeneous is easy to see as well
since, if $c_n > 0$ and $\eta > 0$ we have
\[
V(\eta c', \eta c_n) = (\eta c_n) U(\eta c' / \eta c_n) = \eta (c_n U(c'/c_n)) = \eta V(c', c_n),
\]
while the case where $c_n = 0$ is obvious. Additionally, it is worth noting that
\[
V(c', 1) = U(c'),
\]
so this recovers the original payoff when the num\'eraire's value, $c_n$, is set to 1,
as expected.

\paragraph{Quadratic payoff.} Given the affine case, the next natural question is, can we find the CFMMs corresponding to more complicated
payoff functions? For this case, we will consider a CFMM whose payoff is a concave quadratic. To our knowledge, a CFMM of this form is not known 
in the literature, but the procedure above gives a simple derivation. In particular, we want a CFMM that yields a payoff of
\[
U(c') = -\frac12 (c') ^TA c' + b^Tc' + d,
\]
where $A$ is a strictly positive definite matrix (the case of positive semidefinite matrices is also 
easy to identify,
but requires some additional conditions on $b$ and the nullspace of $A$, which we leave as a simple 
extension). Note that
this is not positive everywhere so the payoff of this CFMM cannot match that of $V'$ everywhere. On 
the other hand, we will see that both are equal within some specific set of cost vectors $c'$.

To start, we will first consider the perspective transformation~\eqref{eq:perspective} of $V'$
to get
\[
V(c', c_n) = \frac{1}{2c_n}(c')^TAc' + b^Tc' + dc_n.
\]
Using~\eqref{eq:linear-offset}, it suffices to consider the simpler function
\[
V(c', c_n) = -\frac{1}{2c_n} (c')^TA c',
\]
as the rest is simply a linear offset of this function, which follows from the previous discussion. 
From~\eqref{eq:main} we have
\[
\psi_V(R', R_n) = \inf_{c_n > 0, c'} ((R')^Tc' + R_nc_n - V(c', c_n))
\]
where $R' \in \reals^{n-1}_+$ are the reserves of the first $n-1$ coins, while $R_n \in \reals_+$ is 
the reserve of the num\'eraire.
To find $\psi_V$ we will first partially minimize over $c'$, using the first order optimality 
conditions, to get
\[
\psi_V(R', R_n) = \inf_{c_n > 0} \left(R_nc_n - \frac{c_n}{2} (R')^TA^{-1}R'\right).
\]
so
\[
\psi_V(R', R_n) = \begin{cases}
	0 & \frac12 (R')^TA^{-1}R' \le R_n\\
	-\infty & \text{otherwise}.
\end{cases}
\]
Finally, adding the linear offset $V(c', c_n) \to V(c', c_n) + a^Tc' + bc_n$, using~\eqref{eq:linear-offset} gives
\[
\psi_V(R', R_n) = \begin{cases}
	0 & \frac12 (R'-a)^TA^{-1}(R'-a) \le R_n - b\\
	-\infty & \text{otherwise},
\end{cases}
\]
as required. Note that, because $U(c')$ is negative and decreasing for large enough $c'$, the 
payoff may not be correctly replicated at all possible price vectors $c'$. In fact
it is not hard to show that once $c'$ is outside of some compact set, the
resulting payoff is always 0, and we leave this as a simple, but interesting, exercise for the 
reader.

\section{Practical applications} \label{sec:practical}
In this section, we outline several practical applications of this general solution method in the more specific case where we have two coins, a traded coin and the num\'eraire.
The first example gives a simple way of reconstructing the well-known constant mean market makers, such as those created
and implemented by Balancer, by attempting to replicate an intuitive payoff function.

We then proceed with a realistic financial product, the covered call.
We present both static replication of the asset payoff at expiry and of the option \emph{price} using the Black-Scholes model. In~\cite{evans2020liquidity} it was shown that the covered call can be statically replicated by a constant mean market maker with dynamic weights.
The replication methodology used here does not require one to update the trading function using an external price oracle. We expect that this will reduce the cost and complexity of implementing these CFMMs in practice.
Finally, we present the example of a perpetual American put option. In this subsection, unlike in the previous subsections, we have that $c_1 \in \reals_+$, $R_1 \in \reals_+$, and
$R_2 \in \reals_+$ are all scalar quantities, where $c_1$ is the price of the asset in question.

\subsection{Balancer}
We can recover some known payoff functions
in a few important cases.
For example, we can ask: what is a CFMM trading function whose payoff is a (concave) power of the price?
In other words, can we find a CFMM whose payoff is:
\[
U(c_1) = c_1^w,
\]
for some $0 < w < 1$?
As is known in the literature (see, \eg,~\cite{AC20, evans2020liquidity, balancer}) we will see that the trading function for Balancer, or that of a constant mean market, is
one such trading function (and, in fact, will be the trading function we recover).

Taking the perspective of $U$ as in~\eqref{eq:perspective}, we have, where $c_2$ is the new variable (the `price of the num\'eraire'), we have
\[
V(c_1, c_2) = c_1^wc_2^{1-w}.
\]
(Note that $V$ is then the weighted geometric mean of $c_1$ and $c_2$ with weights $(w, 1-w)$.)
So, we can recover the trading function by using~\eqref{eq:main},
\[
\psi_V(R_1,R_2) = \inf_{c_2 > 0, c_1} (c_1R_1 + c_2R_2 - V(c_1, c_2)).
\]
This implies that
\begin{equation}\label{eq:geo-mean}
\psi_V(R_1, R_2) = \begin{cases}
 	0 & \left(\frac{R_1}{w}\right)^{w}\left(\frac{R_2}{1-w}\right)^{1-w} \ge 1\\
 	-\infty & \text{otherwise}.
 \end{cases}
\end{equation}
It is also easy to show that
\[
\psi(R_1, R_2) = \left(\frac{R_1}{w}\right)^{w}\left(\frac{R_2}{1-w}\right)^{1-w}
\]
is equivalent to $\psi_V$. We can of course simplify this further by dropping the constant
multiplier $w^{-w}(1-w)^{-(1-w)}$, which yields the usual form for constant mean markets; \ie,
\[
\psi(R_1, R_2) = R_1^{w}R_2^{1-w}.
\]

To show~\eqref{eq:geo-mean}, we consider three separate cases. First, $R_1, R_2 \ge 0$, otherwise $\psi_V$ is unbounded from below. On the other hand, if
\[
\left(\frac{R_1}{w}\right)^{w}\left(\frac{R_2}{1-w}\right)^{1-w} < 1,
\]
then picking $c_1 = tw/R_1$ and $c_2 = t(1-w)/R_2$ for $t \in \reals$ with $t \ge 0$ means that
\[
\psi_V(R_1, R_2) \le c_1R_1 + c_2R_2 - V(c_1, c_2) = t \underbrace{\left(1 - \left(\frac{R_1}{w}\right)^{-w}\left(\frac{R_2}{1-w}\right)^{-(1-w)}\right)}_{< 0} \to -\infty,
\]
as $t \to \infty$. Finally, if
\[
\left(\frac{R_1}{w}\right)^{w}\left(\frac{R_2}{1-w}\right)^{1-w} \ge 1,
\]
then, using the weighted AM-GM inequality we find, for any $c_1, c_2 > 0$,
\[
c_1R_1 + c_2R_2 = wc_1\frac{R_1}{w}+ (1-w)c_2\frac{R_2}{1-w} \ge \left(c_1\frac{R_1}{w}\right)^w\left(c_2\frac{R_2}{1-w}\right)^{1-w} \ge c_1^{w}c_2^{1-w},
\]
which means that $\psi_V(R_1, R_2) \ge 0$. Clearly, equality is achievable by choosing $c_1, c_2 \downto 0$,
yielding~\eqref{eq:geo-mean}.

This proof easily generalizes to the case where $c \in \reals_+^n$ and
\[
V(c) = \prod_{i=1}^n c_i^{w_i},
\]
where $\ones^Tw = 1$ and $w_i > 0$ for $i=1, \dots, n$, with equivalent trading function, for $R \in \reals_+^n$,
\[
\psi(R) = \prod_{i=1}^n \left(\frac{R_i}{w_i}\right)^{w_i}.
\]
(This is just the $n$ coin constant mean market maker, \eg,
Balancer, as is used in practice.)

\subsection{Covered call at expiry}
In the following examples, we assume that we have a `risky' asset with reserve amount $R_1$ and a `risk-free' asset with reserve amount $R_2$.
We seek to find the trading function that corresponds to certain derivative securities.
We will restrict our attention to `covered' instruments whose replication does not require short positions in either asset (negative reserve quantities).
We note that lending markets and offsetting positions have been proposed as solutions for replicating these types of instruments~\cite{evans2020liquidity}, but do not explore this further in this work. 

In this first application, we consider the terminal payoff of a covered call. 
This strategy involves combining a long position in the risky asset with a short position in a call option on the risky asset.
A covered call allows the writer to generate additional income from a long position in exchange for giving up upside for prices above the strike.
At expiry, the payoff of a covered call with strike $K$ at expiry is
\[  
U(c) = c_1 - \max(c_1-K,0)
\]
We take the perspective,
\[
V(c_1,c_2)  = \begin{cases}
	c_1 & c_1 < K \\
  c_2K & c_1 \geq K.
\end{cases}
\]
For $(R_1,R_2) \in \reals^{2}_+$, we have the trading function
\[
\psi_V(R_1,R_2) = \sup_{c} (V(c_1, c_2) - c_1R_1 - c_2R_2)\\
=\min\{f_1(R_1,R_2),f_2(R_1,R_2)\}
\]
where 
\[
f_1(R_1,R_2) = \sup_{c_1 \geq K, c_2>0} (c_1 - c_1R_1 - c_2R_2) = \begin{cases}
	K-KR_1-R_2 = 0 & (R_1,R_2)=(1,0) \\
  +\infty & \text{otherwise}
\end{cases}
\]
and
\[
f_2(R_1,R_2) = \sup_{c_1 < K, c_2>0} (c_2K - c_1R_1 - c_2R_2) = \begin{cases}
		K-KR_1-R_2=0 & (R_1,R_2)=(0,K) \\
 +\infty & \text{otherwise}.
\end{cases}
\]
We therefore have the trading function,
\[
\psi_V(R_1,R_2) = \sup_{c_1 < K, c_2>0} (c_2K - c_1R_1 - c_2R_2) = \begin{cases}
 0 & (R_1,R_2)=(1,0) \\
	0 & (R_1,R_2)=(0,K) \\
	 +\infty & \text{otherwise}.

\end{cases}
\]
In other words, the CFMM only will either hold one unit of the underlying asset or $K$ units of the risk-free asset. When the option is out of the money, $c_1<K$, the CFMM will hold only one unit of the risky asset $(R_1,R_2)=(1,0)$. The CFMM will function equivalently to a limit order to sell one unit of the underlying at $K$. When the option is in the money, the CFMM will therefore hold only $K$ units of the risk free asset, i.e. $(R_1,R_2)=(0,K)$. In either case, the arbitrageur will ensure that the CFMM holdes the lower of of $(0,1)$ and $(0,K)$. This implies that the constant sum curve,

\[
\psi_V(R_1,R_2) =
	K-KR_1-R_2 = 0
\]
will yield the same payoff by allowing the arbitrageur to trade between the underlying and the risk-free asset. This is analogous to the trading rule used in the stop-loss start-gain strategy for replicating an option position~\cite{carr_stop_loss}, but requires full collateralization.

\subsection{Black-Scholes covered call price}
While the previous example gives the terminal payoff of a covered call, it requires full collateralization, which we expect will not be particularly useful in practice. In this example, we instead replicate the \emph{price} of the covered call under the Black-Scholes. We chose this model as a standard example because it is well-studied, but note that our approach could accommodate different pricing models and assumptions. The replication in this section will require less initial capital and apply to the price of the instrument at any time prior to expiry. In this case, given a price $c_1 \ge 0$, we can create a two-coin CFMM whose portfolio value function replicates the Black-Scholes price of a covered call, given by
\[  
U(c_1) = c_1(1-\Phi(d_1))+K\Phi(d_2),
\]
where $\tau>0$ is the time to maturity, $K \geq 0$ is the strike price, $\Phi(\cdot)$ is the standard normal CDF, and
\[
   d_1 = \frac{\log(c_1/K)+ (\sigma^2/2)\tau}{\sigma \sqrt{\tau}}, \qquad d_2 = d_1 - \sigma \sqrt{\tau},
\]
where $\sigma \ge 0$ is the implied volatility.
Here, we assume zero risk-free rate, \ie, $r=0$, but the extension to the case of a positive risk-free rate is immediate. Taking the perspective of $U$
\[
V(c_1, c_2) = c_1-c_1\Phi(d'_1)+Kc_2\Phi(d'_2),
\]
where we have modified the constants to satisfy
\[
\begin{aligned}
d'_1 = \frac{\log(\frac{c_1}{c_2 K})+ (\sigma^2/2)\tau}{\sigma \sqrt{\tau}}, \qquad d'_2 = d'_1 - \sigma \sqrt{\tau}. \nonumber
\end{aligned}
\]
Using~\eqref{eq:main}, we write, for $R, R' \ge 0$,
So, we can recover the trading function by using~\eqref{eq:main},
\[
\psi_V(R_1,R_2) = \sup_{c_2 > 0, c_1} (V(c_1, c_2) - c_1R_1 - c_2R_2).
\]

\begin{figure}
    \centering
    \includegraphics[width=.9\textwidth]{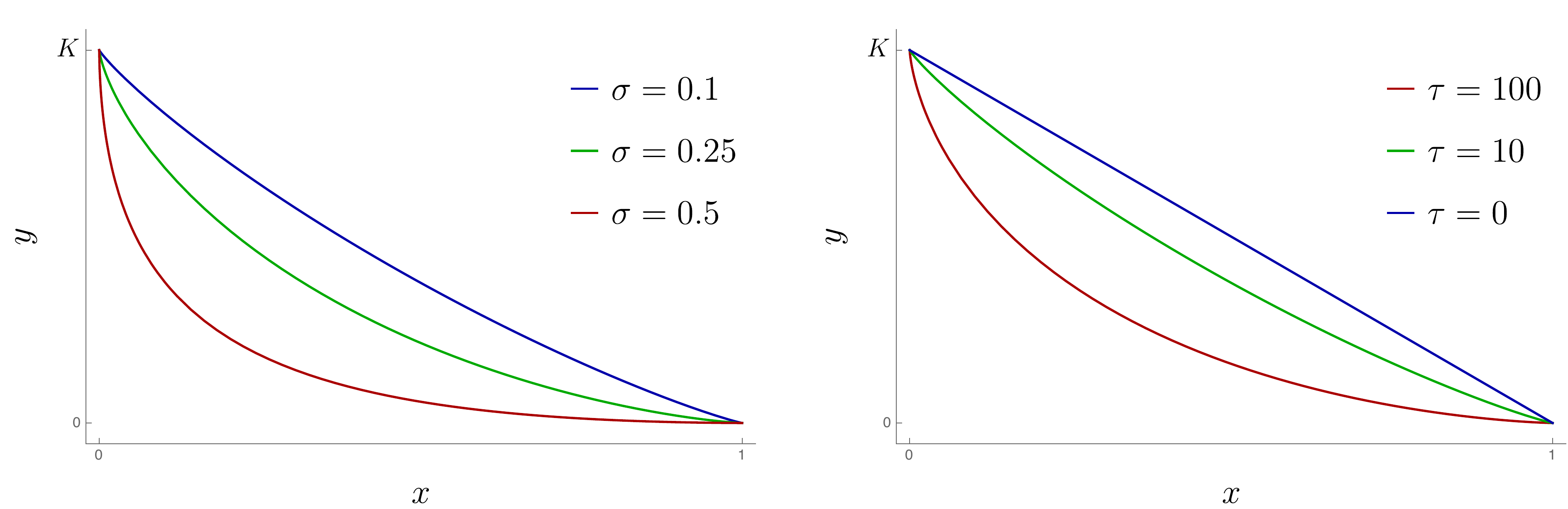}
    \caption{The left figure plots the trading function of the replicating CFMM for a covered call with $\tau=10$ for different values of implied volatility. The right figure shows how the trading function changes with time to maturity for $\sigma=0.1$.}
    \label{fig:covered_call}
\end{figure}

Partially minimizing over $c$, we have the first-order conditions
\[
R_1-1+\Phi(d'_1)=0, \quad d'_1=\Phi^{-1}(1-R_1),\quad c_1=c_2 K h(R_1),
\]
where $h$ is defined as
\[
h(R_1) = e^{\sigma \sqrt{\tau}\Phi^{-1}(1-R_1)-\tau\sigma^2},
\]
for convenience. We can substitute this result back, and, after cancellations, find:
\[
    \psi_V(R_1,R_2) =  \sup_{c_2 > 0} \left(c_2(R_2-K\Phi(\Phi^{-1}(1-R_1)-\sigma \sqrt{\tau})  \right).
\]
It is then immediate that:
\[
\psi_V(R_1, R_2) = \begin{cases}
	0 & R_2-K\Phi(\Phi^{-1}(1-R_1)-\sigma \sqrt{\tau}) \le 0\\
	+\infty & \text{otherwise}.
\end{cases}
\]
We plot some examples in Figure~\ref{fig:covered_call}. When the price of the risky asset and the time to matury are both strictly positive, the covered call payoff will require more capital to replicate a covered call closer to maturity. This can be seen in both the formula for $V'(c)$ and visually in Figure~\ref{fig:covered_call}. If one were to update the trading function over time as the option neared maturity, each update would require additional capital. The difference over time is determined by `theta' which captures the time-decay of the option's price. The no-fee CFMM will not capture gains from theta decay and as such will fail to offer self-financing replication. We conjecture that fees may restore the ability of the LP to profit from theta decay when replicating such positions, as demonstrated in a simpler case in~\cite{AEC21}. We discuss theta decay further in~\S\ref{app:delta-hedging}, but do not resolve this question directly in this work.

\subsection{Perpetual American put option price}
We consider a payoff of $K-P(c)$, where $P(c)$ is the Black-Scholes price of a perpetual put option struck at $K$. From~\cite[Chapter~6]{shreve2004stochastic}, this payoff is 
\[
U(c_1)  = \begin{cases}
	c_1 & c_1 \leq L \\
  K - (K-L)(\frac{c_1}{L})^{-\frac{2r}{\sigma^2}} & c_1>L.
\end{cases}
\]
where $L=\frac{2r}{2r+\sigma^2}K$, $\sigma$ is the volatility of the risky-asset and $r$ is the risk-free rate.
Taking the perspective, 
\[
V(c_1,c_2)  = \begin{cases}
	c_1 & c_1 \leq L \\
  c_2K - c_2^{\frac{2r+\sigma^2}{\sigma^2}}(K-L)(\frac{c_1}{L})^{-\frac{2r}{\sigma^2}} & c_1>L.
\end{cases}
\]
We have the trading function
\[
\psi_V(R_1,R_2) = \sup_{c_2 > 0, c_1} (V(c_1, c_2) - c_1R_1 - c_2R_2)=\min\{f_1(R_1,R_2),f_2(R_1,R_2)\}
\]
where 
\[
f_1(R_1,R_2) = \sup_{c_1 \leq L, c_2>0} (c_1 - c_1R_1 - c_2R_2) = \begin{cases}
	L-LR_1-R_2 = 0 & (R_1,R_2)=(1,0) \\
  +\infty & \text{otherwise}
\end{cases}
\]
and
\[
f_2(R_1,R_2) = \sup_{c_2 > 0, c_1>L} (c_2K - c_2^{\frac{2r+\sigma^2}{\sigma^2}}(K-L)(\frac{c_1}{L})^{-\frac{2r}{\sigma^2}} - c_1R_1 - c_2R_2)
\]
The first-order condition is
\[
c_1=c_2 \left(\frac{R_1\sigma^2}{2r}\frac{L^{-\frac{2r}{\sigma^2}}}{K-L}\right)^{-\frac{\sigma^2}{2r+\sigma^2}}
\]
Substituting this back and after some cancellations we get


\[
\psi_2(R_1,R_2) = \sup_{c_2 > 0} (c_2(K - R_2 - KR_1^{\frac{2r}{2r+\sigma^2}}))
= \begin{cases}
	 0 & K - R_2 - KR_1^{\frac{2r}{2r+\sigma^2}} \leq 0\\
  +\infty & \text{otherwise}
\end{cases}
\]
Noting that $f_1(1,0)=f_2(1,0)=0$, we can simply use $f_2$ as our trading function. Unlike the previous example, the perpetual American put does not require one to update the curve or contribute additional capital to continue the replication.

\begin{figure}
    \centering
    \includegraphics[width=.9\textwidth]{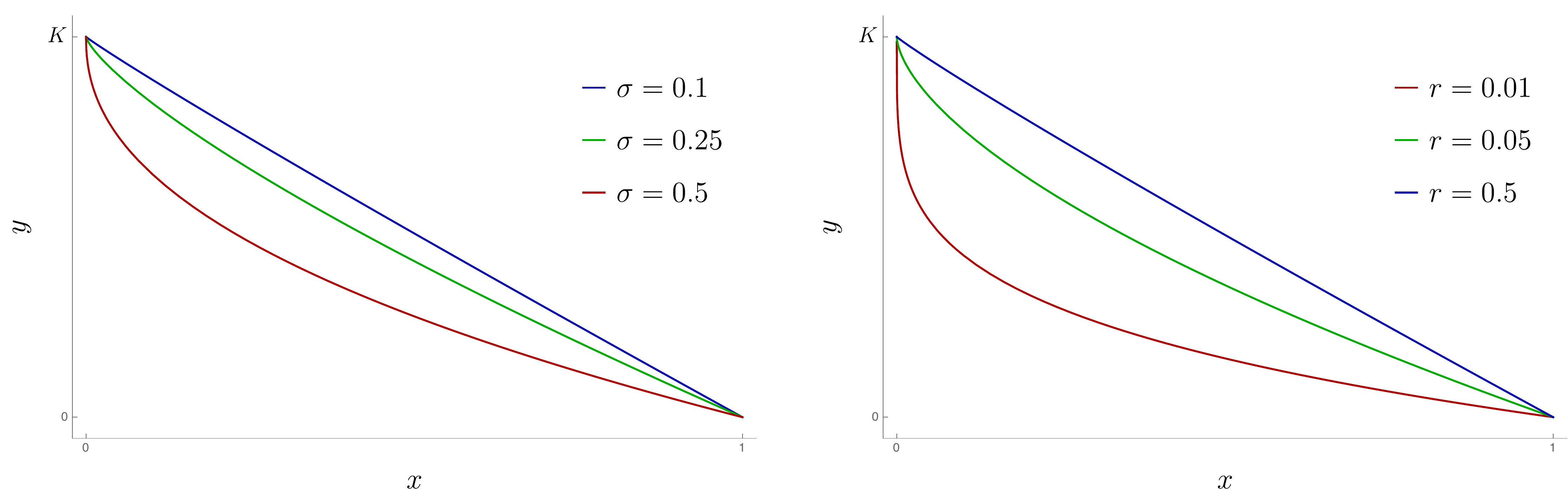}
    \caption{The left figure plots the trading function of the replicating CFMM for a perpetual American put with $r=0.1$ for different values of implied volatility. The right figure plots the trading function for different values of $r$ when $\sigma=0.25$.}
    \label{fig:perpetual_put}
\end{figure}

\section{Conclusion} \label{sec:conclusion}

We demonstrated an equivalence between CFMM trading functions and consistent (\ie, concave, 
nonnegative, nondecreasing, 1-homogeneous) payoff functions. Our methodology relies only on 
basic tools from convex analysis and can produce the appropriate trading function for 
replicating a number of theoretically and practically interesting payoffs. However, we also 
point to cases where replication requires additional initial capital due to arbitrage costs. 
Finally, the results presented in this article may have implications for the design of 
existing CFMMs and indeed qualitatively matches some results in previous work. In 
particular, the result of~\cite{AEC20} that lower-curvature CFMMs are more suitable to lower-
volatility assets appears to be confirmed in the discussion of options in \ref{sec:practical}, 
wherein the trading functions have lower curvature for lower implied volatility (for a visual 
illustration, see figures~\ref{fig:covered_call} and~\ref{fig:perpetual_put}).

\paragraph{Future work.} There are several interesting directions for future research. For 
example, determining whether fees can mitigate 
arbitrage losses, allowing replication without additional capital, would be a very useful 
result. We conjecture that such a 
result would generalize the framework of~\cite{AEC21} from constant-mean marker makers to 
arbitrary CFMMs. As previously discussed, another possibility is, when the replication of a 
given payoff is impossible, how closely one can approximate the payoff with a CFMM. Another 
area for research involves extending the results to include convex payoffs as well as
positions that require leverage to replicate. Convex instruments may require the ability to 
establish short positions in CFMMs shares. Similarly, levered instruments require one to facilitate 
lending secured by the value of CFMM shares. A model for secure lending and borrowing of CFMM 
shares would therefore expand the range of payoffs one can replicate with a CFMM.


\bibliographystyle{alpha}
\bibliography{bib}

\appendix

\section{Delta hedging} \label{app:delta-hedging}

As in section \ref{sec:practical}, we have that $c_1 \in \reals_+$, $R_1 \in \reals_+$, and
$R_2 \in \reals_+$ are all scalar quantities, where $c_1$ is the price of the asset in question. We seek to construct a trading function such that $R_1=R_1(c_1)$ for all $c_1$. We define,
\[
p(R_1)=R_1^{-1}(c_1),
\]
which is the marginal price of the traded coin. Recalling that $R_2$ can be thought of as an implicit function of $R_1$, we have 
\[
\frac{dR_2}{dR_1}=-p(R_1).
\]
Therefore,
\[
\int p(R_1)dR = -R_2
\]
will give a family of trading functions with the desired property. 

\paragraph{Hedging a covered call.} Extending the example in section \ref{sec:practical}, we now look at delta hedging a covered call. In this case, we hold $\frac{d}{d c} V(c)$ units of the risky asset, i.e. $R_1(c)=1-\Phi(d_1)$. We therefore have,
\[
\int Kh(R_1) = -R_2,
\]
where $h$ is defined as before where $h$ is defined as $h(R) = e^{\sigma \sqrt{\tau}\Phi^{-1}(1-R)-\tau\sigma^2}$. In this case, we have the trading function
\[
\psi(R_1,R_2) = k-\frac{1}{2}K+K\Phi(\Phi^{-1}(1-x)-\sigma \sqrt{\tau})-R_2=0
\]
where k is an arbitrary constant. As before, $k$ will control where the hedge will fail as the CFMM runs out of reserves. An appropriate choice wil be to select $k$ such that the intial value of reserves  match the initial price of the covered call. In this case, we have
\[
c_1R_1+R_2=c_1(1-\Phi(d_1))+K\Phi(d_2).
\]
Substituting the values for $R_1(c)=1-\Phi(d_1)$ and $R_2=k-\frac{1}{2}K+K\Phi(\Phi^{-1}(1-x)-\sigma \sqrt{\tau})$ and solving for $k$, one obtains $k=\frac{K}{2}$. Substituting this value recovers the trading function we derived in \ref{sec:practical}.

\paragraph{Hedging a log contract.} We consider the case of delta-hedging a short position in a log contract. As noted in \cite{Neuberger74}, delta-hedging a contract paying the natural logarithm of the futures price will replicate a variance swap. In this case, we seek a trading function for which $R_1(c)=\frac{1}{c_1}$. In other words, to achieve the desired hedge, we seek a CFMM that holds one unit of asset 2 worth of asset 1. Noting that, 
\[
p(R_1)=\frac{1}{R_1}
\]
we get
\[
\int p(R_1)dR_1 =k-\ln{R_1}=R_2,
\]
where $k$ is an arbitrary constant. Any trading function of the form
\begin{equation} \label{eq:LMM}
\psi(R_1,R_2) = k-\ln{R_1}-R_2=0
\end{equation}
will achieve the desired hedge insofar as $k \geq \ln{R_1}$ or $c_1 \geq e^{-k}$, as the CFMM will otherwise run out of the reserves required to continue hedging.

\begin{figure}
    \centering
    \includegraphics[width=0.75\textwidth]{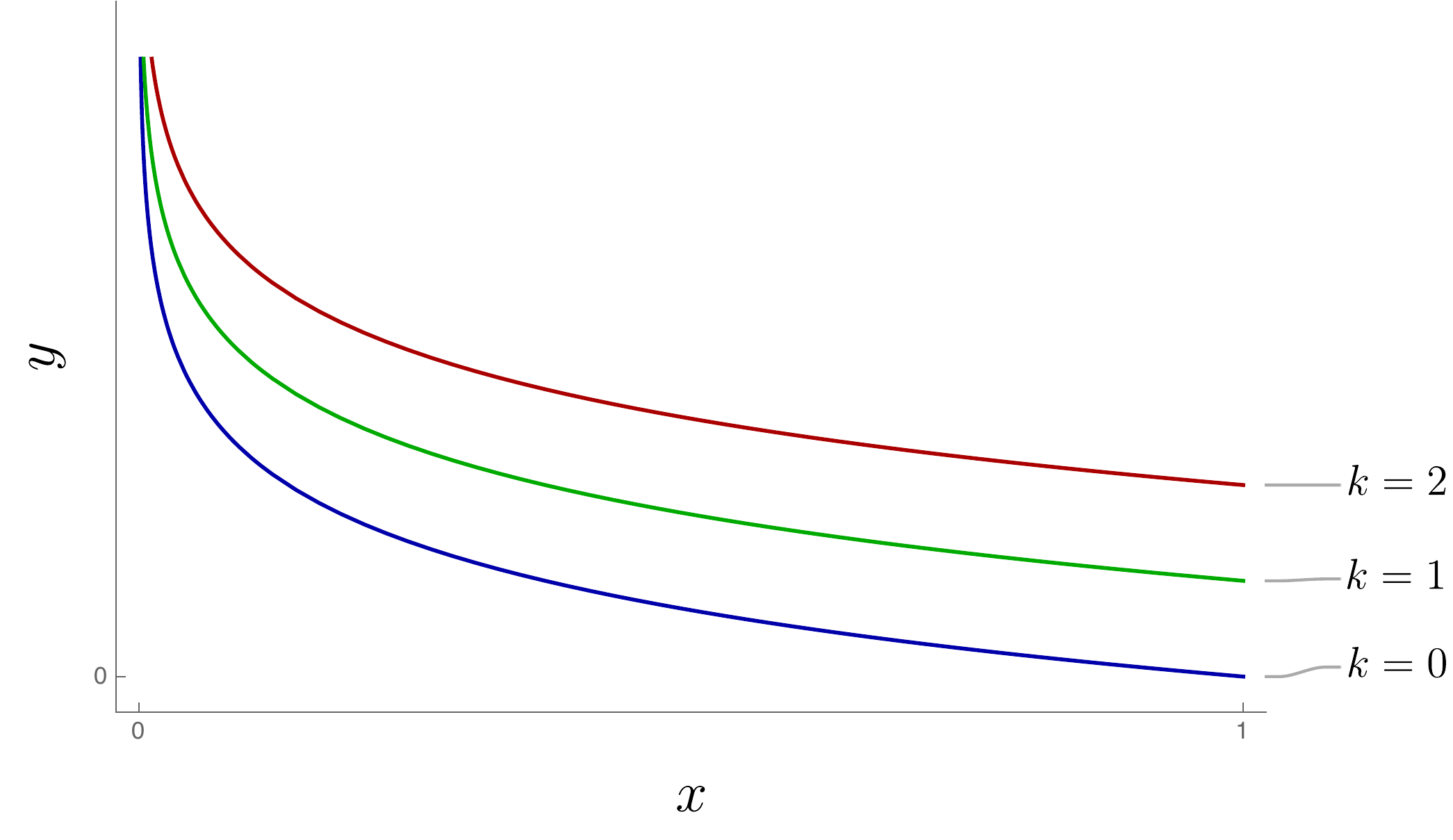}
    \caption{Delta-hedging CFMM for the log contract with different parameter choices for the constant $k$}
    \label{fig:LMM}
\end{figure}

\paragraph{Path dependence and arbitrage loss.} Suppose the price of the asset at time $t$ is $c_1(t)$ and consider delta-hedging a short position in the log contract by holding $\frac{1}{c_1(t)}$ units of the asset under zero transaction costs. The PNL of this strategy over a discrete period $[t,t+1]$ is $\frac{1}{c_1(t)}(c_1(t+1)-c_1(t))$. When continuously rebalancing over $[0,T]$, we have PNL $\int_0^T \frac{1}{c_1(t)} dc_1(s)$. For simplicity of illustration, suppose $c_1(t)$ follows a geometric Brownian motion with stochastic differential
\[
dc_1(t)=\sigma dW(t),
\]
where $W(t)$ is a standard Brownian motion. One can check that the expected PNL of the delta-hedging strategy is zero. Now, we contrast this with delta-hedging with the CFMM we recovered in \eqref{eq:LMM}. One can check that this CFMM has payoff
\[
V(c_1)= k+\ln{c_1}.
\]
The expected PNL of this strategy is therefore, 
\[
\Expect[V(c_1(T))-V(c_1(0))]=-\frac{\sigma^2}{2}T
\]
In other words, implementing the delta-hedge through a no-fee CFMM instead of continual rebalancing under no transaction costs will result in a supermartingale. This observation is analogous to the result in \cite{AC20,evans2020liquidity} that the portfolio value of a G3M or constant-mean market maker is a supermartingale under the risk-neutral measure due to arbitrage losses.

More generally, a no-fee CFMM has payoff $V(c)=R_1c + R_2$, which is always path-independent. In contrast, the equivalent delta-hedging strategies continually-rebalanced at no cost are path-dependent. When delta-hedging a convex strategy, one's portfolio will be short gamma and long theta~\cite{shreve2004stochastic}. The equivalent CFMM does not benefit from positive theta due to arbitrage, resulting in the supermartingale behavior. In other words, delta hedging a convex claim with a CFMM with the appropriate concave payoff will underperform the equivalent delta hedge rebalanced under no transaction costs. We conjecture that a result similar to that of~\cite{AEC21} may allow one to arbitrarily approximate unconstrained delta-hedging strategies with CFMMs by taking the directional limit as the fee approaches zero, but do not pursue this direction further in this paper.

\end{document}